\documentclass[%
 reprint,
 amsmath,amssymb,
 aps,
]{revtex4-2}
\usepackage[utf8]{inputenc}
\usepackage{graphicx}  
\usepackage{dcolumn}   
\usepackage{bm}        
\usepackage{amssymb,amsmath,diagbox,adjustbox}   
\usepackage{color}
\usepackage{appendix}

\usepackage{graphicx}
\usepackage{dcolumn}
\usepackage{bm}

\begin{document}


\title{Extracting Cosmological Information from Lightcone Data: A Comparison of CNNs and Summary-Statistic–Based Approaches}

\author{Zhiwei Min}
 \affiliation{School of Physics and Astronomy, Sun Yat-Sen University, Zhuhai 519082, China}
 \author{Xu Xiao}
 \affiliation{School of Physics and Astronomy, Sun Yat-Sen University, Zhuhai 519082, China}
    \author{Zhujun Jiang}
 \affiliation{School of Physics and Astronomy, Sun Yat-Sen University, Zhuhai 519082, China}
 \author{Fenfen Yin}
 \affiliation{School of Physics and Electronic Information Engineering, Tongren University, Tongren 554300, China}
\author{Jiacheng Ding}
 \affiliation{Shanghai Astronomical Observatory, Chinese Academy of Sciences, Shanghai 200030, P. R. China}
 \author{Liang Xiao}
 \affiliation{School of Physics and Astronomy, Sun Yat-Sen University, Zhuhai 519082, China}
  \author{Haitao Miao}
 \affiliation{National Astronomical Observatories, Chinese Academy of Sciences, 20A Datun Road, Beijing 100101, China}
 \author{Shupei Chen}
\affiliation{Pengcheng Laboratory, Nanshan District, Shenzhen 518000, China}
\author{Qiufan Lin}
\affiliation{Pengcheng Laboratory, Nanshan District, Shenzhen 518000, China}
\author{Yang Wang}
 \affiliation{Pengcheng Laboratory, Nanshan District, Shenzhen 518000, China}
\author{Le Zhang}
 \email{zhangle7@mail.sysu.edu.cn}
 \affiliation{School of Physics and Astronomy, Sun Yat-Sen University, Zhuhai 519082, China}
 \author{Xiao-Dong Li}
 \email{lixiaod25@mail.sysu.edu.cn}
 \affiliation{School of Physics and Astronomy, Sun Yat-Sen University, Zhuhai 519082, China}

\date{\today}

\begin{abstract}
Lightcone observations are the natural data format of galaxy surveys, but their evolving geometry breaks the translational symmetry assumed by standard convolutional neural networks (CNNs). In particular, applying CNNs to 3D gridded lightcone data implicitly treats the line-of-sight direction as translationally invariant, despite encoding cosmic time evolution. We propose a simple alternative  (CNN+2D) that divides the lightcone into redshift slices, projects each onto a HEALPix sphere, and analyzes them with a 2D CNN. Using \texttt{AbacusSummit} halo lightcone mocks ($0.3<z<0.8$, $40^\circ\times40^\circ$), we compare this approach with fully connected  networks (FC) applied to different summary statistics, including spherical harmonic coefficients ($a_{\ell m}$), wavelet scattering transform (WST) coefficients, and the angular two-point correlation function (2PCF), along with standard 2PCF likelihood and Fisher forecasts. We find that multiple statistics beyond CNNs can achieve competitive performance: FC networks combined with $a_{\ell m}$ and especially WST significantly outperform 2PCF-based methods, with FC+WST yielding the best overall parameter constraints across cosmologies. Meanwhile, for a fiducial cosmology with multiple realizations, the CNN+2D approach achieves the smallest statistical uncertainties. These results demonstrate that both learned features and carefully constructed summary statistics can effectively extract cosmological information from lightcone data, providing flexible and robust analysis strategies for upcoming surveys such as DESI, Euclid, and CSST.

\end{abstract}
\maketitle


\section{Introduction}
\label{sec:intro}

The large-scale structure (LSS) of the Universe encodes essential information about its composition and evolution \citep{Peacock1999, Peebles2020}. The observed distribution of galaxies and dark matter arises from primordial fluctuations and is shaped by cosmological parameters \citep{Planck2018, Springel2005}. By comparing observational data with theoretical predictions and numerical simulations, we can place stringent constraints on the underlying physics of the Universe \citep{Tegmark2004, Weinberg2013}.

The LSS exhibits a characteristic web-like pattern of clusters, filaments, walls, and voids \citep{Bond1996, Hahn2007}, formed through the combined effects of gravitational instability and cosmic expansion \citep{Davis1985, Peebles1980}. To connect these structures with cosmological models, the galaxy distribution is commonly compressed into two-point statistics such as the power spectrum and correlation function \citep{Hamilton1993, Feldman1994}. These statistics have been highly successful in providing constraints on key physical effects, including baryon acoustic oscillations (BAO) \citep{Eisenstein2005}, the Alcock–Paczynski distortion \citep{Alcock1979}, and redshift-space distortions (RSD) \citep{Kaiser1987}.

However, two-point statistics are complete only for Gaussian density fields. The late-time cosmic web is highly non-Gaussian due to nonlinear gravitational evolution, and significant cosmological information resides in the non-Gaussian features such as phase correlations and higher-order clustering \citep{Bernardeau2002}. To access this information, higher-order statistics such as the three-point correlation function (3PCF) and its Fourier analog, the bispectrum, have been developed and applied to galaxy surveys \citep{Takada:2002qq}. These can be extended to even higher orders (e.g., four-point correlation function or trispectrum), but the computational cost and complexity of interpretation grow rapidly with order, making a complete description of the non-Gaussian field practically infeasible \citep{Slepian:2015qza}.

Recent advances in machine learning have provided new ways to exploit the rich structure of cosmological datasets \citep{LeCun2015, Goodfellow2016}. In particular, deep learning methods, especially convolutional neural networks (CNNs), have shown strong potential for extracting cosmological information directly from simulated or observed matter fields. Unlike traditional summary statistics that require manual selection of which information to compress, neural networks can learn optimal representations directly from the data, naturally capturing information from all orders of statistics. CNN have been used to predict cosmological parameters from N-body density fields \citep{Ribli2019, Fluri2019}, to recover parameters such as the matter density $\Omega_m$, fluctuation amplitude $\sigma_8$, and spectral index $n_s$~\citep{Schmelzle2017, Lucie-Smith2020}. More recently,~\citet{Min:2024dgd} utilized the \texttt{Quijote} LH$\nu w$ simulations to infer parameters in a 7-dimensional cosmological parameter space, while other studies have applied CNN to infer cosmology from mock galaxy catalogs~\citep{He2019, Ntampaka2020,Saez-Casares:2026hwv}. These studies demonstrate that neural networks can capture information beyond traditional summary statistics, offering a promising complement to established techniques. 

With the rapid expansion of CNN applications in LSS studies, applying these methods to galaxy redshift survey data--such as lightcone observations--provides a more realistic setting for cosmological analyses. \citet{Ntampaka_2020} demonstrated one of the first applications of 3D CNNs to discretized galaxy survey data, converting mock catalogs from the \texttt{AbacusSummit} simulations into data cubes of size $550 \times 550 \times 220 \, (h^{-1}\mathrm{Mpc})^3$, where the $220 \, h^{-1}\mathrm{Mpc}$ dimension corresponds to the LoS. More recently, \citet{Hwang:2023oob} developed a 3D CNN framework to constrain cosmological parameters 
($\Omega_m$, $\sigma_8$, the dark energy equation-of-state today $w_0$, and the derived clustering parameter $S_8$) 
from simulated dark matter halo lightcones. These lightcones span $40^\circ \times 40^\circ$ on the sky over the redshift range $0.3 < z < 0.8$, and are binned into $64^3$ voxels for CNN processing.

Directly applying standard CNNs to lightcone data faces several challenges: full lightcone volumes mapped onto 3D grids are extremely large, making 3D CNNs computationally costly; the line-of-sight direction traces cosmic time evolution, so the statistical properties vary with redshift, breaking the translational invariance assumed by conventional CNNs; and in the transverse direction, the data lie on the curved celestial sphere, whereas standard CNNs are designed for flat Euclidean grids. To overcome these issues, we introduce a scheme that enables the use of efficient 2D CNNs. The method decomposes the 3D lightcone into narrow redshift slices, within which cosmic evolution is slow enough to approximate the sky as statistically stationary, restoring translation invariance along each slice. Each slice is then projected onto a HEALPix spherical map \citep{gorski2005healpix}, whose equal-area, hierarchical pixelization preserves the spherical geometry while providing a structured grid compatible with convolutional operations. This combination produces a representation aligned with the symmetries assumed by CNNs, allowing efficient extraction of physically meaningful cosmological features while avoiding the computational cost of full 3D convolutions.

To assess the performance of our framework, we apply the 2D CNN pipeline to \texttt{AbacusSummit} halo lightcone mocks ($0.3<z<0.8$, $40^\circ \times 40^\circ$), where the 3D lightcone is divided into narrow redshift slices and projected onto HEALPix spheres. We benchmark this approach against fully connected  networks (FC) trained on various summary statistics, including $a_{\ell m}$, WST coefficients, and the 2PCF, as well as against standard 2PCF likelihood analyses and Fisher forecasts. We find that several alternative statistics yield competitive constraints: FC networks with $a_{\ell m}$, and especially with WST, outperform the 2PCF-based method, with FC+WST providing the tightest bounds across cosmologies. For a fiducial cosmology with multiple realizations, the method CNN+2D consistently achieves the smallest statistical uncertainties, highlighting its ability to efficiently extract cosmological information.

This paper is structured as follows. Section~\ref{sect:data} describes the AbacusSummit data used in this work, details the preprocessing steps applied to incorporate observational effects, and outlines the strategy employed to render the lightcone data compatible with CNN-based analysis. Section~\ref{sec:method} introduces the machine learning models applied to 2D HEALPix maps. Section \ref{sec:statistics} presents a summary-statistic-based fully connected network. It details the process of preparing the input statistics, including the spherical harmonic coefficients $a_{\ell m}$, the angular two-point correlation function, and the WST coefficients.
 Section~\ref{sec:results} reports the results from all models and compares their performance. Finally, Section~\ref{sec:conclusion} summarizes our findings and suggests directions for future work.


\section{Data}
\label{sect:data}
To carry out our analysis, we utilize the \texttt{AbacusSummit} simulation suite \citep{maksimova2021abacussummit}, which consists of a set of large, high-precision cosmological N-body simulations performed using the \texttt{ABACUS} N-body code \citep{garrison2019high,garrison2021abacus}. These simulations provide detailed halo catalogs and lightcone outputs. This allows us to generate observationally-motivated mock halo lightcones for validating our framework. The suite was specifically designed to satisfy and surpass the cosmological simulation requirements of the DESI survey \citep{levi2013desi,levi2019dark}.

\texttt{AbacusSummit} includes over 150 simulations, covering 97 different cosmologies and containing a total of approximately 60 trillion particles. For the present analysis, we focus exclusively on the \texttt{base} configuration boxes. Each of these boxes contains $6912^3$ particles within a $(2~h^{-1}\mathrm{Gpc})^3$ volume, corresponding to a particle mass of $2\times10^9~h^{-1}\mathrm{M}_\odot$ and a force softening of $7.2~h^{-1}$ kpc. Halo identification in the \texttt{AbacusSummit} suite is performed using the \texttt{COMPASO} halo finder, which is based on a specialized spherical-overdensity algorithm \citep{hadzhiyska2022compaso}.

We specifically selected 52 simulations (\texttt{AbacusSummit base c\{130-181\} ph000}), ranging from \texttt{c130} to \texttt{c181}, which share the same initial random seed.  This set provides broad coverage of the eight-dimensional cosmological parameter space. The base cosmology used in these simulations is consistent with the Planck 2018 results \citep{aghanim2018planck}.

For our analysis, the set of parameters employed as data labels is $\{\Omega_b, \Omega_m, h, 10^9 A_s, n_s, \sigma_8\}$. We do not attempt to constrain the parameters of dynamical dark energy ($w_0$, $w_a$), because the majority of the cosmologies are set exactly at the base values ($w_0 = -1$, $w_a = 0$).

\subsection{Halo Selection}
\label{ssec:lightcone}

The \texttt{AbacusSummit} halo light-cone catalogs are constructed from the corner of the main simulation box and two periodic copies, seamlessly attached to provide continuous coverage \citep{hadzhiyska2022halo}. For this work, we select halos with $\mathrm{RA}/\mathrm{DEC} \in (0, 40^\circ)$ and redshift $0.3 < z < 0.8$, producing a complete halo catalog over this footprint and redshift range.

The halo mass function is highly sensitive to cosmological parameters \citep{tinker2008toward}. To preserve cosmological information while minimizing biases from resolution effects, we apply a minimum halo mass cut of $M_{\rm min} = 2\times 10^{11}~h^{-1}\mathrm{M}_\odot$, corresponding to at least 200 particles per halo. This low-mass selection yields a high number density of halos, ranging from  $7.98\times 10^{-3}$ to $10.8\times 10^{-3}~(h/\mathrm{Mpc})^3$. This range is comparable to the expected galaxy number densities for fourth-generation redshift surveys \citep{CSST2019, CSST2026, 2025PhRvD.112h3515A}, allowing us to test our framework under a sampling density relevant to future observations. 

\subsection{Observational Effects}
\label{ssec:obs_effects}

To make the simulated data more realistic and consistent with observational analysis pipelines, we apply two critical post-processing steps to the halo catalogs.

\subsubsection{Redshift-Space Distortions}
\label{sssec:rsd}

To mimic the observational effect along LoS, halo positions are modified as
\begin{equation}\label{eq:rsd}
\bm{s}_{\parallel} = \bm{r}_\parallel + \frac{\bm{v} \cdot \hat{z}}{a H(a)} \hat{z}\,,
\end{equation}
where $\bm{r}_\parallel$ and $\bm{s}_\parallel$ are the original and RSD-distorted positions along the LoS, respectively. Here, $\hat{z}$ is the unit vector along the LoS, $\bm{v}$ is the peculiar velocity of the halo, $a$ is the scale factor, and $H(a)$ is the Hubble parameter. The perpendicular coordinates remain unchanged.

\subsubsection{Mapping to a Fiducial Cosmology}
\label{sssec:fiducial}

Observational data are interpreted under a fiducial cosmology, so all mock catalogs are transformed to match the fiducial background (\texttt{c000}). The transformation is given by
\begin{equation}\label{eq:fiducial}
s_\perp = s_\perp^0 \frac{d_A^{f}(z)}{d_A(z)}\,, \quad
s_\parallel = s_\parallel^0 \frac{H(z)}{H^{f}(z)}\,,
\end{equation}
where $s_\perp^0$ and $s_\parallel^0$ are the original comoving coordinates perpendicular and parallel to LoS, $d_A(z)$ and $H(z)$ are the angular diameter distance and Hubble parameter at redshift $z$, and the superscript $f$ denotes the fiducial cosmology. This procedure converts redshifts to comoving distances consistently with real observations \citep{alam2017clustering}.

\subsection{Preprocessing Lightcone Slices}
\label{sssec:2dmap}

In our scheme, the 3D lightcone data are first transformed into a representation compatible with 2D convolutions. In the following, we describe the details of this preprocessing pipeline, which consists of two key steps: redshift binning and HEALPix projection.

\textbf{Redshift binning.} We utilize the nine narrow redshift slices provided by the \texttt{AbacusSummit} lightcone outputs over $0.3 < z < 0.8$: $[0.275, 0.325]$, $[0.325, 0.376]$, $[0.376, 0.425]$, $[0.425, 0.475]$, $[0.475, 0.538]$, $[0.538, 0.613]$, $[0.613, 0.688]$, $[0.688, 0.763]$, and $[0.763, 0.838]$. Within each thin slice, cosmic evolution and selection effects are approximately constant, effectively restoring translation symmetry along LoS and facilitating subsequent 2D CNN analysis.

\textbf{HEALPix projection.} For each redshift slice, halos are projected onto a HEALPix pixelized sphere, which provides an equal-area, hierarchical grid that preserves the sky's spherical geometry. We adopt a resolution of $N_{\mathrm{side}}=512$, yielding $N_{\mathrm{pix}} = 12 N_{\mathrm{side}}^2 \approx 3.1\times 10^6$ pixels. Each pixel records the number of halos it contains, forming a 2D density map for the slice. This procedure produces a data cube of shape $9 \times 512 \times 512$ for each cosmological model, where the first dimension indexes the ordered redshift slices, enabling a standard 2D CNN to process the data while respecting both spherical geometry and approximate translational symmetry within each slice. 

The HEALPix maps are generated using the \texttt{healpy} Python package \citep{2005ApJ...622..759G,Zonca2019}. For each slice, we employ the \texttt{CartesianProj} tool to map the spherical HEALPix map onto a $512\times512$ Cartesian grid covering the $40^\circ \times 40^\circ$ sky region. Each Cartesian pixel is assigned by tracing its center coordinates $(\theta,\phi)$ back to the corresponding HEALPix pixel, ensuring an equal-area projection from the sphere to the plane.

\begin{figure}
    \centering
    \includegraphics[width=0.8\linewidth]{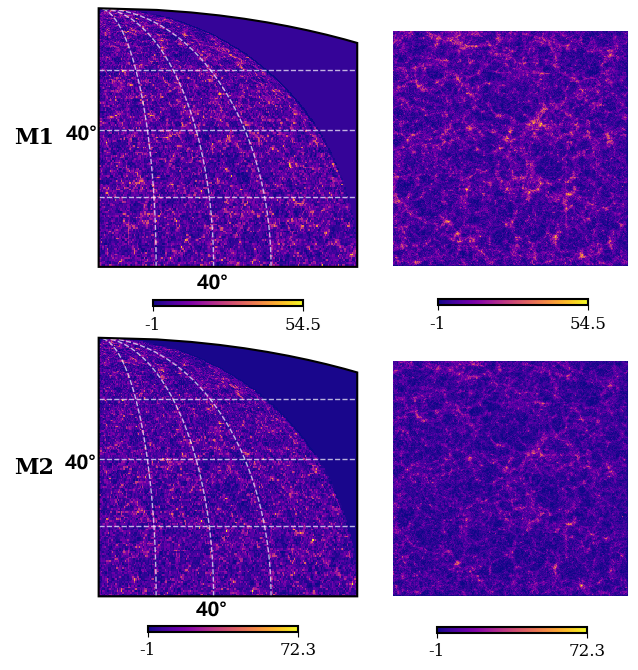}
    \caption{HEALPix maps and the corresponding projected 2D density fields for two cosmological models (M1 and M2) in the redshift bin $0.475 \le z \le 0.538$.  
\textit{Left:} full-sky simulation covering RA and DEC from $0^\circ$ to $90^\circ$.  
\textit{Right:} zoomed-in region with RA and DEC both from $0^\circ$ to $40^\circ$.  
The color bar indicates the overdensity.
}
\label{fig:HEALPix_2dproj}
\end{figure}

Figure~\ref{fig:HEALPix_2dproj} presents the 1D HEALPix density maps for two cosmological models, M1 and M2, within the redshift bin $0.475 \le z \le 0.538$, shown in two projections. The left panel displays a Mollweide projection covering RA $0$–$90^\circ$ and DEC $0$–$90^\circ$, while the right panel shows a Cartesian projection of a zoomed region (RA $0$–$40^\circ$, DEC $0$–$40^\circ$). The cosmological parameters are: 
M1: $\{\Omega_b = 0.0473, \Omega_m = 0.3210, h = 0.6618, 10^9 A_s = 2.52, n_s = 0.9303, w_0 = -1, w_a = 0, \sigma_8 = 0.8825\}$;  
M2: $\{\Omega_b = 0.0528, \Omega_m = 0.3578, h = 0.6472, 10^9 A_s = 1.65, n_s = 0.9252, w_0 = -1, w_a = 0, \sigma_8 = 0.7297\}$. The color scale represents the matter overdensity, defined as $\delta = \rho/\bar{\rho} - 1$. Because both simulations share identical initial conditions, the observed differences in the evolved structures are entirely due to variations in the cosmological parameters. 

Note that, although the $40^\circ \times 40^\circ$ sky patch does not strictly satisfy the flat-sky approximation, the Cartesian projection used here is purely a mathematical mapping. This representation is intended to facilitate the subsequent application of 2D CNNs, and it preserves the relative spatial relationships of halos within each slice while remaining compatible with standard convolution operations.

\section{CNN model for lightcones}
\label{sec:method}

We assess two types of machine learning models designed to predict the six cosmological parameters $\{\Omega_b, \Omega_m, h, A_s, n_s, \sigma_8\}$ from different representations of the halo lightcone data. 

The first architecture is a two-dimensional convolutional neural network (2D CNN) applied to the HEALPix-projected 2D density maps. In this network, convolutional layers extract spatial features, which are then processed by FC layers to infer the target parameters. The second architecture exclusively employs FC layers, which take as input vectorized summaries: PCA-compressed $a_{\ell m}$ coefficients, 2PCF measurements, and PCA-compressed WST coefficients. In this section, we detail the 2D CNN architecture; the summary-statistic-based FC model will be introduced in the following section \ref{sec:statistics}.

\begin{figure}[htpb]
    \centering
    \includegraphics[width=0.95\linewidth]{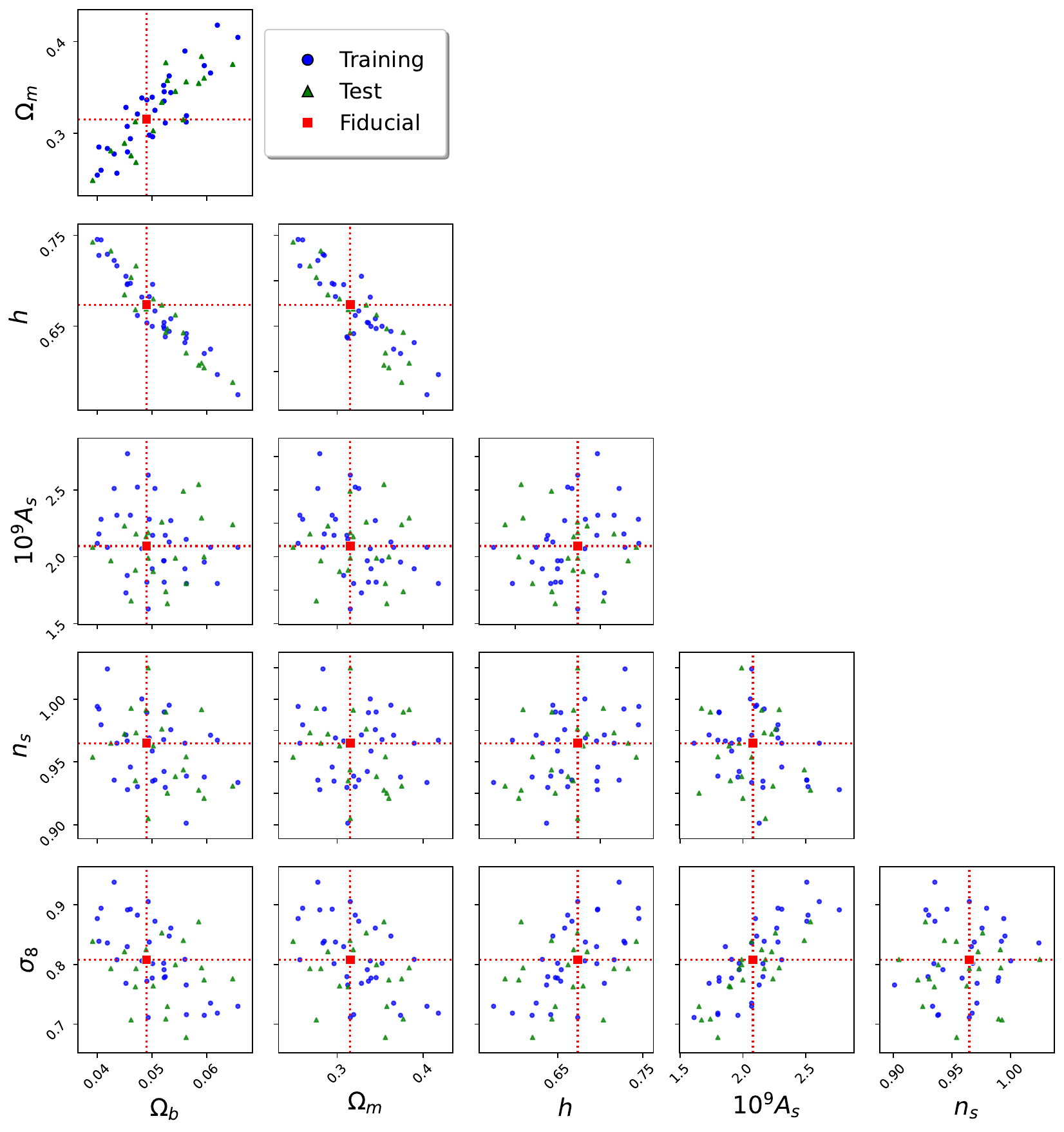}
    \caption{Distribution of cosmological parameters for the 52 simulations from \texttt{AbacusSummit base c130-181 ph000}. The red square marks the fiducial base cosmology (\texttt{c000}). Blue points indicate the 32 training samples, and green points represent the 20 test samples.}
    \label{fig:para_dis}
\end{figure}

The 52 cosmological simulations are split into 32 training models and 20 test models. Distributions of the 6 parameters for the training set, test set, and fiducial values are shown in Fig.~\ref{fig:para_dis}. All networks are trained using the Adam optimizer \citep{Kingma2014AdamAM} with an initial learning rate of $10^{-4}$ and a batch size of 4. The loss function is defined as the mean squared error (MSE) between the predicted and true cosmological parameters:
\begin{equation}
\mathcal{L_{\rm loss}} = \frac{1}{N} \sum_{i=1}^{N} \sum_{\alpha = 1}^6 \left( \hat{\theta}_\alpha^{i} - \theta_{\alpha}^{i} \right)^2,
\end{equation}
where $\hat{\theta}^i_{\alpha}$ and $\theta^i_{\alpha}$ represent the predicted and true values, respectively, of the $\alpha$-th cosmological parameter for the $i$-th test sample. Here, $N$ is the number of test samples, and the summation over $\alpha$ from 1 to 6 corresponds to the six cosmological parameters in our parameter set.

To quantify the predictive accuracy of the models, we use the coefficient of determination ($R^2$) and the root-mean-square error (RMSE), defined as

\begin{equation}\label{eq:metrics}
\begin{aligned}
R^2 &= 1 - \frac{\displaystyle\sum_{i=1}^{n} (\theta_i - \hat{\theta}_i)^2}{\displaystyle\sum_{i=1}^{n} (\theta_i - \bar{\theta})^2}, \\
\mathrm{RMSE} &= \sqrt{\frac{1}{n} \displaystyle\sum_{i=1}^{n} (\theta_i - \hat{\theta}_i)^2}\,.
\end{aligned}
\end{equation}
These metrics together characterize both the goodness of fit and the typical prediction error of each model. The $R^2$ value measures the proportion of variance in the true parameters that is explained by the model's predictions, with a value closer to 1 indicating better agreement between predictions and true values, while RMSE provides an absolute measure of the typical prediction error in the natural units of each parameter.

\subsection{2D CNN for Projected Density Fields of Lightcones}

To directly exploit the spatial structure of the lightcone data, we employ a purpose-designed 2D CNN applied to the HEALPix-projected 2D density maps for cosmological parameter inference.

The 2D CNN architecture (Fig.~\ref{fig:2dconvnet}) is designed to extract cosmological information from the HEALPix-projected 2D density maps. The input, consisting of 9 redshift bins of $512\times512$ pixels, is processed through four convolutional blocks with increasing filters $(32,64,128,256)$, capturing multi-scale spatial features. An adaptive average pooling layer reduces the feature maps to a fixed size of $256\times16^2$, which is then flattened into a 1D vector of $65,536$ elements. This feature vector is fed into a fully connected block that outputs the 6 cosmological parameters. The following subsections describe each component in detail, beginning with the convolutional block.

\subsubsection*{Convolutional Block}
Each block contains:
\begin{itemize}
    \item \textbf{Convolution:} $3\times 3$ kernel, stride 1, zero padding to preserve spatial dimensions.
    \item \textbf{Batch normalization:} Normalizes outputs to accelerate training and improve stability.
    \item \textbf{ReLU activation:} Applies $f(x) = \max(0,x)$ element-wise for non-linearity and sparse representation.
    \item \textbf{Max-pooling:} $2\times 2$ kernel, stride 2, reducing spatial size by half and providing translation invariance.
\end{itemize}

\subsubsection*{Fully Connected Block}
Each FC block includes:
\begin{itemize}
    \item \textbf{Input layer:} 1D vector of length $n$, adjustable for different inputs (e.g., $256\times16^2$ for HEALPix maps, $9\times26$ for 2PCF, $9\times185$ for $a_{\ell m}$, $64\times60$ for WST).
    \item \textbf{Dense layer:} 512 neurons with ReLU activation.
    \item \textbf{Dropout:} Rate 0.3 to prevent overfitting.
    \item \textbf{Output layer:} 6 neurons for cosmological parameter predictions ($\Omega_b, \Omega_m, h, A_s, n_s, \sigma_8$).
\end{itemize}

\begin{figure}[htpb]
    \centering
    \includegraphics[width=0.95\linewidth]{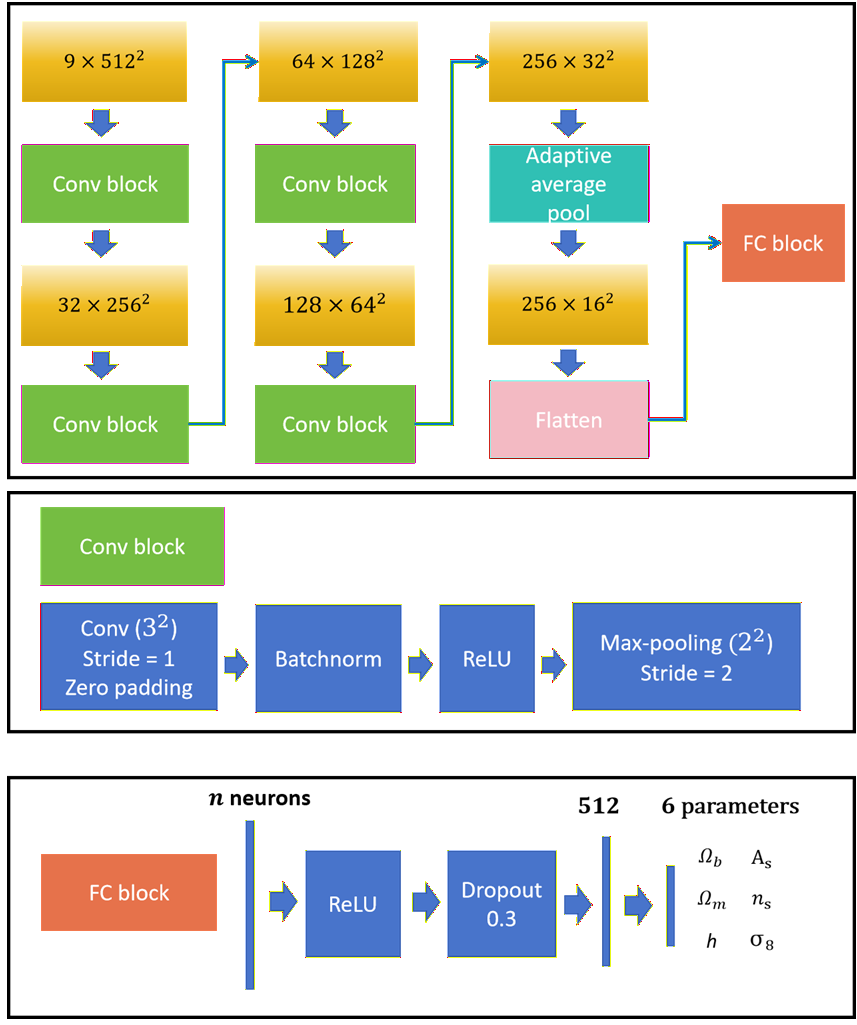}
    \caption{2D CNN architecture for HEALPix-projected density fields ($9 \times 512^2$, 9 redshift bins). The network contains four convolutional blocks (convolution + batch normalization + $2\times2$ max-pooling), which progressively reduce the spatial resolution by a factor of two, followed by an adaptive pooling layer and a FC regressor for cosmological parameter estimation. For models using summary statistics (2PCF, $a_{\ell m}$, and WST), the convolutional feature-extraction stage is removed and only the same FC regressor is retained. The input dimension $n$ of the FC block is adjusted according to match the dimensionality of each statistic.}
    \label{fig:2dconvnet}
\end{figure}

\section{Summary-Statistic-Based FC Network}
\label{sec:statistics}

To evaluate the performance of our CNN+2D approach, we consider several summary statistics coupled with a FC network, including the $a_{\ell m}$ coefficients, the 2PCF, and the WST coefficients, followed by a comparison of their performance. 

\subsection{Spherical Harmonic Coefficients}
\label{sssec:alm}

We extract the spherical harmonic coefficients ($a_{\ell m}$), from each HEALPix-projected redshift slice and compress them using principal component analysis (PCA). As a natural basis for functions on the sphere, any square-integrable field $f(\theta,\phi)$ can be expanded as
\begin{equation}
    f(\theta,\phi) = \sum_{\ell=0}^{\infty} \sum_{m=-\ell}^{\ell} a_{\ell m} Y_{\ell m}(\theta,\phi)\,,
\end{equation}
with coefficients obtained via
\begin{equation}
    \label{eq:alm}
    a_{\ell m} = \frac{4\pi}{N_{\mathrm{pix}}} \sum_{p=1}^{N_{\mathrm{pix}}} Y^*_{\ell m}(\theta_p,\phi_p) f(\theta_p,\phi_p)\,,
\end{equation}
where the sum runs over all $N_{\mathrm{pix}}$ HEALPix pixels. We adopt a maximum multipole of $\ell_{\mathrm{max}} = 256$. The angular power spectrum 
\begin{equation}
C_\ell = \frac{1}{2\ell+1} \sum_{m=-\ell}^{\ell} |a_{\ell m}|^2\,,
\end{equation}
is thus derived from the spherical harmonic coefficients $a_{\ell m}$ of each HEALPix map.

The resulting coefficients are high-dimensional. Exploiting the Hermitian property $a_{\ell,-m} = (-1)^m a_{\ell m}^*$, only $m \ge 0$ coefficients are retained, giving 33,153 independent complex values. Separating real and imaginary parts produces a 66,306-dimensional feature vector per redshift slice. To reduce dimensionality while preserving most of the information, we apply PCA to the concatenated feature vectors from all 52 cosmologies and 9 redshift slices (468 samples in total), retaining 185 principal components that explain 99\% of the variance. This compresses each slice to a 185-dimensional representation, and the resulting features are reshaped into a dataset of shape $(52,\,9\times185)$, which serves as the input to a fully connected neural network.

\subsection{Angular Two-Point Correlation Function}
\label{sssec:2pcf}

For comparison with traditional clustering statistics, we compute the angular two-point correlation function (2PCF), $\omega(\theta)$. In terms of the projected overdensity field $\delta(\hat{\mathbf{n}})$ on the sky, it is defined as
\begin{equation}
\omega(\theta) = \left\langle \delta(\hat{\mathbf{n}}_1)\,\delta(\hat{\mathbf{n}}_2) \right\rangle,
\end{equation}
where the average is taken over all pairs of directions $\hat{\mathbf{n}}_1$ and $\hat{\mathbf{n}}_2$ separated by an angle $\theta$.

We estimate $\omega(\theta)$ using the public code \texttt{CUTE} with the Landy \& Szalay estimator \citep{landy1993bias},
\begin{equation}
\omega(\theta) =
\frac{DD(\theta) - 2DR(\theta) + RR(\theta)}{RR(\theta)},
\end{equation}
where $DD$, $DR$, and $RR$ denote the normalized counts of data–data, data–random, and random–random pairs. The random catalog contains ten times more points than the halo catalog to reduce shot noise, with positions uniformly distributed in RA $[0,2\pi)$ and in $\sin(\mathrm{DEC})$ $[-1,1]$, ensuring isotropic, equal-area coverage. This geometry-corrected set is essential for unbiased estimates.

The correlation function is measured separately for each of the nine redshift slices over the angular range $0.17^\circ < \theta < 8.5^\circ$, using 26 linearly spaced bins. To capture the redshift evolution, the measurements from all slices are concatenated into a single feature vector, resulting in a dataset of shape $(52, 9\times26)$ that serves as the input to a fully connected neural network.

\subsection{Wavelet Scattering Transform Coefficients}
\label{sssec:wst}

The wavelet scattering transform (WST) was originally introduced in the context of signal processing and computer vision \citep{mallat2012group, bruna2013invariant}, providing a framework to capture hierarchical statistical properties of an input field. In the WST, a field $I(\bm{x})$ undergoes successive nonlinear operations: convolutions with wavelets and modulus transformations. For an oriented wavelet $\Psi_{j_1, l_1}(\bm{x})$ probing scale $j_1$ and angle $l_1$, the first step is

\begin{equation}
I'(\bm{x}) = \left| I(\bm{x}) \otimes \Psi_{j_1, l_1}(\bm{x}) \right|,
\label{eq:wst}
\end{equation}

where $\otimes$ denotes convolution. Averaging over the spatial domain produces WST coefficients $S_n$, which describe the field's characteristics. Repeated application of wavelets at multiple scales and orientations forms a scattering network, with coefficients up to order $n=2$ given by

\begin{equation}
\begin{aligned}
S_0 &= \langle |I(\bm{x})| \rangle, \\
S_1(j_1,l_1) &= \langle | I(\bm{x}) \otimes \Psi_{j_1, l_1}(\bm{x}) | \rangle, \\
S_2(j_2,l_2,j_1,l_1) &= \big\langle \big| | I(\bm{x}) \otimes \Psi_{j_1,l_1}(\bm{x}) | \otimes \Psi_{j_2,l_2}(\bm{x}) \big| \big\rangle,
\end{aligned}
\label{eq:wst_coeff}
\end{equation}

where $\langle \cdot \rangle$ denotes averaging over the spatial domain. The wavelets are generated by dilations and rotations of a mother wavelet; here, we use solid harmonics modulated by a Gaussian:

\begin{equation}
\Psi_l^m(\bm{x}) = \frac{1}{(2\pi)^{3/2}} e^{-|\bm{x}|^2/2\sigma^2} |\bm{x}|^l Y_l^m\Big(\frac{\bm{x}}{|\bm{x}|}\Big),
\end{equation}

where $Y_l^m$ are the spherical harmonics and the Gaussian width is set to $\sigma = 0.25$ in field pixels.

The oriented wavelet filter $\Psi_{j,l}(\bm{x})$ used in Eqs.~(\ref{eq:wst}) and (\ref{eq:wst_coeff}) is obtained by averaging (or summing) over the azimuthal index $m$ of the dilated and rotated mother wavelets. Specifically, for scale $j$ and angular index $l$, the filter is constructed as:
\begin{equation}
\Psi_{j,l}(\bm{x}) = \sum_{m=-l}^{l} \lambda_j^{-d}\, \Psi_l^m\!\left(\lambda_j^{-1}\bm{x}\right),
\end{equation}
where $\lambda_j = 2^{j}$ is the dilation factor, and $d$ is the spatial dimension. This operation produces a real-valued, oriented wavelet sensitive to angular frequency $l$.

Unlike the previous statistics, which are computed separately for each redshift slice, the WST is applied directly to the full lightcone volume without binning along the redshift direction. Instead, we preserve the continuous redshift evolution and divide the lightcone only in the angular (RA–DEC) directions. This strategy allows the WST to capture non-Gaussian structures while retaining the full line-of-sight information.

The preparation of WST coefficients from the \texttt{AbacusSummit} halo lightcones proceeds as follows:

\begin{itemize}

\item \textbf{3D grid construction.} 
Halos in the redshift range $0.3<z<0.8$ are gridded into a 3D density field using the Cloud-in-Cell (CIC) scheme, in order to construct the data cube needed for computing the 3D WST coefficients. The grid has dimensions $512\times512\times64$, where the first two axes correspond to RA and DEC and the third axis corresponds to redshift $z$. This representation preserves the full three-dimensional structure of the lightcone.

\item \textbf{Sub-region division.} 
To control the computational cost of the WST calculation while retaining the complete redshift information, the $512\times512\times64$ grid is divided into $8\times8=64$ non-overlapping sub-regions in the RA–DEC plane. Each sub-region therefore has dimensions $64\times64\times64$, preserving the entire redshift information. 

\item \textbf{WST coefficient computation.} 
For each $64\times64\times64$ sub-volume, we compute the WST coefficients independently. We adopt $J=6$ dyadic scales and $L=6$ angular orientations, resulting in 784 WST coefficients per sub-volume (excluding the zeroth-order coefficient $S_0$).

\item \textbf{Data aggregation.} 
With 52 cosmological models and 64 sub-regions per model, this procedure produces a coefficient matrix of shape $(52\times64,\,784)$, corresponding to 3,328 samples.

\item \textbf{PCA dimensionality reduction.} 
The raw WST coefficients exhibit strong correlations. We therefore apply PCA and retain components that explain 99\% of the variance. This compresses the feature vector for each sub-volume to 60 dimensions. The final PCA-compressed dataset has shape $(52,\,64\times60)$ and is used as the input to a fully connected neural network.

\end{itemize}

\subsection{FC Networks for Summary Statistics}

Based on the cosmological sensitivity of the summary statistics (Appendix~\ref{ssec:stat_comparison}), we evaluate their performance within a unified inference framework.

For the three input representations---$a_{\ell m}$ coefficients, 2PCF, and WST coefficients---we employ a FC network. To enable a fair comparison with CNN+2D, the FC network uses the same regressor architecture (the FC block in Fig.~\ref{fig:2dconvnet}), with the convolutional feature-extraction stage removed.

The input layer is adjusted to match the dimensionality of each statistic ($9\times185$ for $a_{\ell m}$, $9\times26$ for 2PCF, and $64 \times 60$ for WST). All other training hyperparameters are kept identical to those used for the CNN+2D model. With this setup, the FC networks take the summary statistics directly as input vectors and learn the mapping from these compressed representations to the cosmological parameters. This design allows us to isolate the impact of different summary statistics on parameter inference while keeping the regression architecture fixed.

\section{Results}
\label{sec:results}

In this section, we present the results of our cosmological parameter inference experiments, focusing on a systematic comparison between the ``CNN+2D'' approach applied to HEALPix-projected density fields (Sect.~\ref{sssec:2dmap}) and several alternative models based on summary statistics.

Specifically, we consider inference pipelines in which different summary statistics are combined with a FC network. These include:
(i) PCA-compressed $a_{\ell m}$ coefficients (``FC+alm''; see Sect.~\ref{sssec:alm}),
(ii) the two-point correlation function (``FC+2PCF''; see Sect.~\ref{sssec:2pcf}), and
(iii) PCA-compressed WST coefficients (``FC+WST''; see Sect.~\ref{sssec:wst}).

We first evaluate the overall performance of the four models on the test dataset (Sect.~\ref{ssec:overall_perf}), and then assess their constraining power using a fiducial cosmology test set with multiple realizations at fixed parameters not included in the training data. Finally, we compare the deep learning results with traditional 2PCF likelihood analysis and Fisher forecasts.

\subsection{Comparison of Model Performance on the Test Set}
\label{ssec:overall_perf}

\begin{figure}
    \centering
    \includegraphics[width=0.9\linewidth]{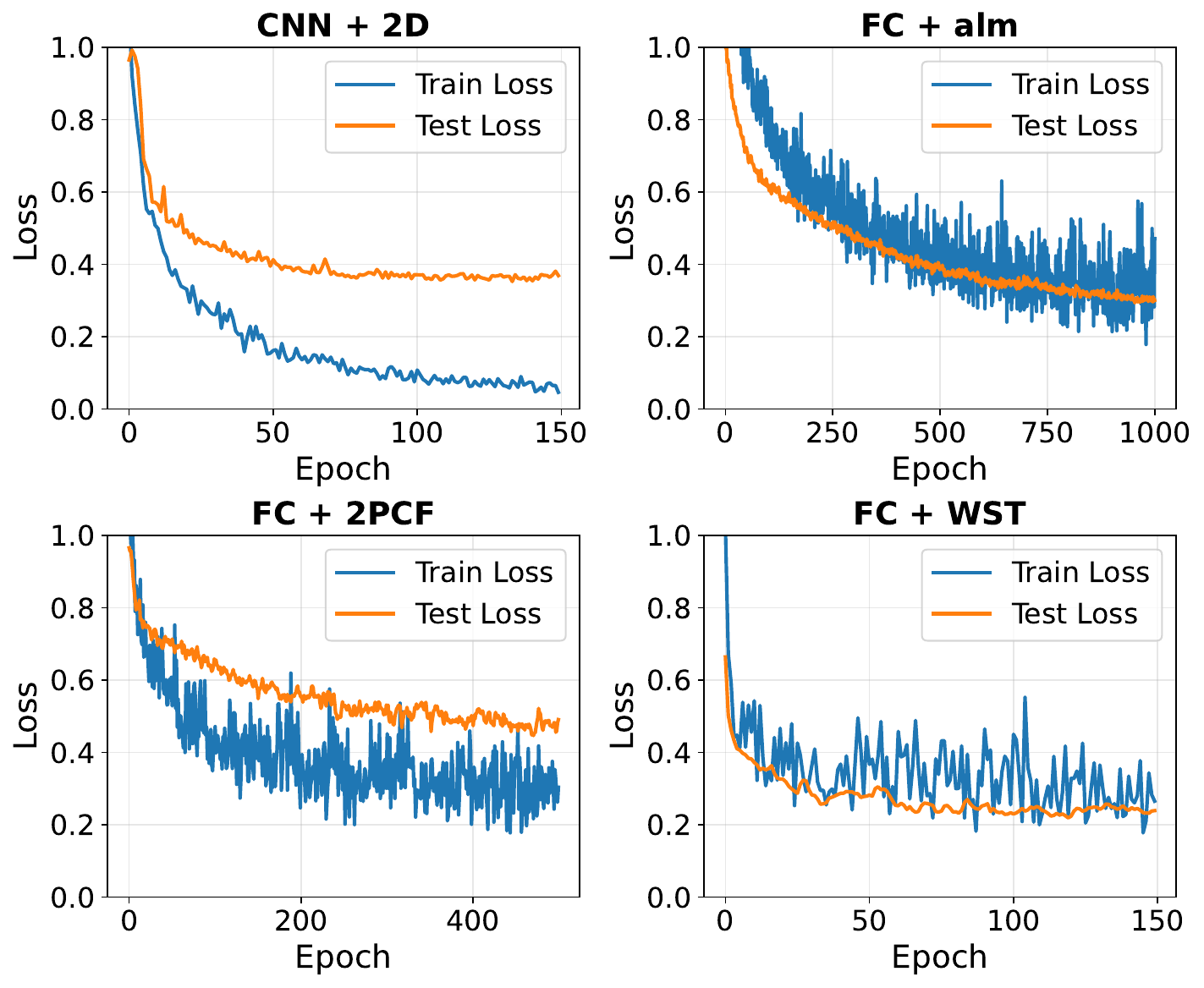}
    \caption{Training and test loss curves for the four models. The blue line represents the training loss, and the orange line represents the test loss.}
    \label{fig:loss}
\end{figure}

We train the CNN+2D, FC+alm, FC+2PCF, and FC+WST networks on 32 cosmological models and evaluate their performance on the test set of 20 cosmological models. Figure~\ref{fig:loss} shows the training and test loss curves for all four deep learning models. In all cases, the loss decreases rapidly during the early stages of training and gradually approaches convergence after $\sim$100–1000 epochs. The optimal network parameters are selected based on the minimum test loss. 

The minimum test losses achieved are $\mathcal{L}_{\rm loss}=0.353$ for CNN+2D, $0.294$ for FC+alm, $0.430$ for FC+2PCF, and $0.205$ for FC+WST. These results suggest that the FC+WST model yields the most accurate parameter predictions among the methods considered, while FC+2PCF shows comparatively weaker performance. The larger gap between training and test loss in FC+2PCF indicates a tendency toward overfitting, whereas the FC+alm and FC+WST models exhibit more consistent training and test behavior, suggesting better generalization.

\begin{figure*}
    \centering
    \includegraphics[width=0.7\linewidth]{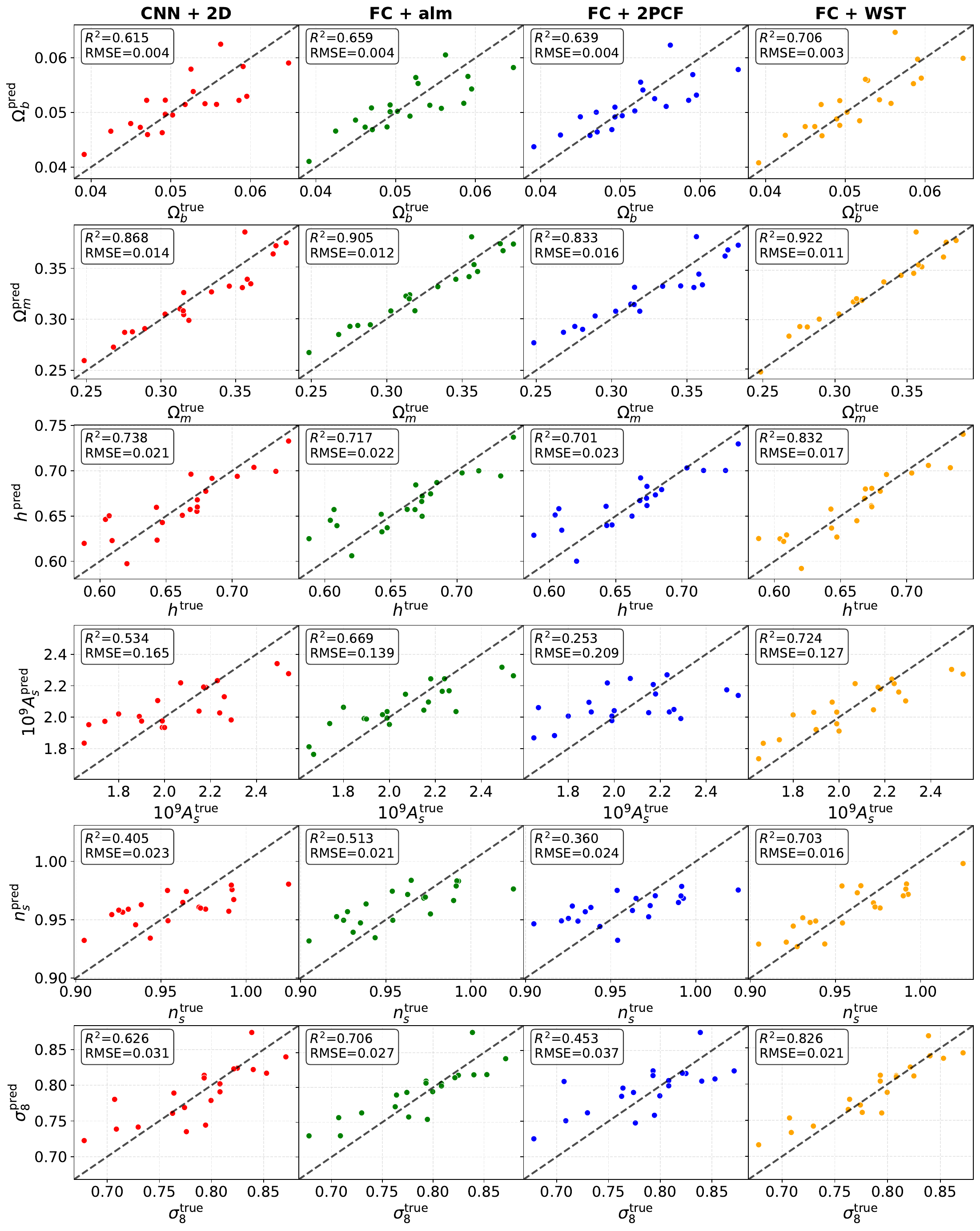}
    \caption{Comparison between the true and predicted values of the cosmological parameters $\Omega_b$, $\Omega_m$, $h$, $A_s$, $n_s$, and $\sigma_8$ obtained using four different methods. The black dashed line indicates the line of perfect agreement. For each panel, the corresponding $R^2$ and RMSE values are displayed in the upper-left corner.}
    \label{fig:pred}
\end{figure*}

Figure~\ref{fig:pred} shows a one-to-one comparison between the true and predicted cosmological parameters for the 20 test models, with $R^2$ and RMSE (defined in Eq.~\ref{eq:metrics}) quantifying the predictive accuracy. The results indicate noticeable differences in performance among the models.

For the baryon density parameter $\Omega_b$, all four methods yield relatively similar RMSE values around 0.003–0.004, while the coefficient of determination $R^2$ varies from 0.615 for CNN+2D up to 0.706 for FC+WST. Among the approaches, the WST-based fully connected network consistently shows the highest $R^2$ and lowest RMSE, suggesting slightly better predictive precision, whereas CNN+2D and the 2PCF-based methods show moderate performance.

The matter density parameter $\Omega_m$ is generally better constrained across all methods. Here, the CNN+2D achieves $R^2 = 0.868$ and RMSE $= 0.014$, while FC+alm and FC+WST improve further to $R^2 = 0.905$ and $0.922$ with RMSE $= 0.012$ and $0.011$, respectively. The 2PCF-based network performs slightly worse ($R^2 = 0.833$, RMSE $= 0.016$). Overall, FC+WST delivers the highest predictive accuracy for $\Omega_m$.

For the Hubble parameter $h$, CNN+2D provides $R^2 = 0.738$ and RMSE $= 0.021$, comparable to FC+alm and FC+2PCF, which exhibit slightly lower $R^2$ and higher RMSE. FC+WST stands out with $R^2 = 0.832$ and RMSE $= 0.017$, indicating improved prediction accuracy relative to other methods.

Predictions for the scalar amplitude $A_s$ show larger discrepancies among methods. CNN+2D and FC+2PCF exhibit lower $R^2$ values of 0.534 and 0.253, with correspondingly higher RMSE (0.165 and 0.209), whereas FC+alm and FC+WST achieve better performance, with FC+WST reaching the best result ($R^2 = 0.724$, RMSE $= 0.127$). This suggests that higher-order statistics captured by the WST help better constrain the amplitude of scalar fluctuations.
For the scalar spectral index $n_s$, FC+WST again provides the most accurate predictions ($R^2 = 0.703$, RMSE $= 0.016$), outperforming CNN+2D ($R^2 = 0.405$, RMSE $= 0.023$), FC+alm ($R^2 = 0.513$, RMSE $= 0.021$), and FC+2PCF ($R^2 = 0.360$, RMSE $= 0.024$).

Finally, for the matter fluctuation amplitude $\sigma_8$, a clear hierarchy in performance emerges. CNN+2D and FC+2PCF yield moderate $R^2$ values of 0.626 and 0.453, respectively, while FC+alm improves to $R^2 = 0.706$. FC+WST achieves the best results with $R^2 = 0.826$ and RMSE $= 0.021$, highlighting its advantage in capturing nonlinear information relevant for structure growth.

In summary, across all cosmological parameters, the WST-based FC network consistently achieves the highest $R^2$ and lowest RMSE, indicating superior overall predictive capability. CNN+2D generally provides moderate performance, particularly for $\Omega_m$ and $h$, while the traditional 2PCF and spherical harmonic-based statistics show intermediate results depending on the parameter. These comparisons demonstrate the relative strengths of different statistical summaries in capturing the relevant information for cosmological parameter inference.

\subsection{Fiducial Cosmology Test}
\label{sssec:single_cosmo}
To assess the statistical constraining power of the trained models, we perform a fiducial cosmology test using the fiducial model (\texttt{c000}). We use 21 independent realizations not included in the training set and generate predictions from each model, with all input representations constructed following Sect.~\ref{sect:data} and Sect \ref{sec:statistics}. The scatter of the predictions reflects the combined effects of cosmic variance and intrinsic model uncertainties. Although limited, the number of realizations is fixed by the \texttt{AbacusSummit} simulation suite, and all methods are evaluated on the same set, enabling a fair comparison. The resulting constraints are further compared with Fisher forecasts and a traditional likelihood analysis.

The 2PCF likelihood function is defined as:

\begin{equation}
\label{eq:likelihood}
\log \mathcal{L}(\theta) = -\frac{1}{2} (d - m(\theta))^T {\mathbf C}^{-1} (d - m(\theta)),
\end{equation}
where $d$ is the 2PCF vector measured from the fiducial realizations, $m(\theta)$ is the model prediction at parameters $\theta$, and ${\mathbf C}$ is the covariance matrix estimated via jackknife resampling (Appendix~\ref{app:jackknife}). The model prediction $m(\theta)$ is obtained from a 2PCF emulator based on Gaussian Process regression (GPR)~\citep{2021arXiv210205497B}, trained on the 52 cosmological models. Its accuracy is validated using leave-one-out cross-validation (Appendix~\ref{app:loo}).

In addition to the likelihood analysis, we compute the Fisher matrix to estimate the expected parameter uncertainties from 2PCF:
\begin{equation}
\label{eq:fisher}
F_{\alpha\beta} = \frac{\partial O_i}{\partial \theta_\alpha} C_{ij}^{-1} \frac{\partial O_j}{\partial \theta_\beta},
\end{equation}
where $O_i$ denotes the 2PCF measurements, and the derivatives are evaluated numerically using the GPR emulator. The marginalized $1\sigma$ uncertainty on each parameter is given by $\sigma_{\theta_\alpha} = \sqrt{(F^{-1})_{\alpha\alpha}}$, which provides an estimate of the achievable constraints within the Fisher approximation.

\begin{figure}[htpb]
    \centering
    \includegraphics[width=0.95\linewidth]{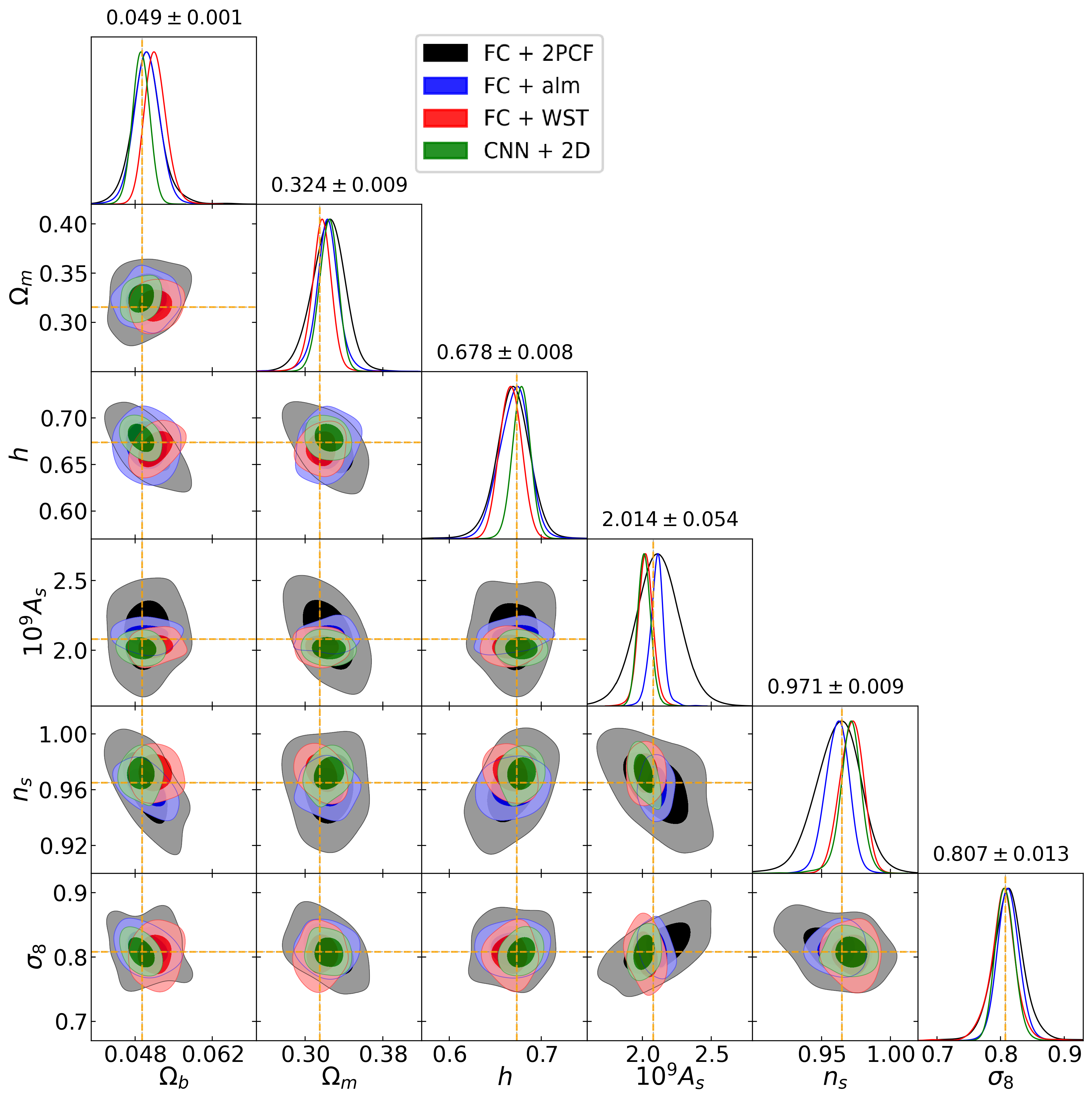}
    \caption{Comparison of the four models in the fiducial cosmology test. The simulations (\texttt{c000}) are generated at fixed fiducial cosmological parameters with different realizations. Diagonal panels show the marginalized 1D posterior distributions, while off-diagonal panels show the 2D $1\sigma$ and $2\sigma$ confidence contours. Orange dashed lines indicate the true parameter values. The $1\sigma$ constraints labeled above the 1D marginals correspond to the CNN+2D model.}
    \label{fig:contour}
\end{figure}

\begin{table*}
\centering
\renewcommand{\arraystretch}{1.5}
\caption{Marginalized posterior constraints (mean $\pm 1\sigma$ uncertainty) for the six cosmological parameters recovered using the four models on the fiducial cosmology. For comparison, the Fisher-matrix forecast and the full likelihood analysis results for the 2PCF are also shown.}
\label{tab:1d_constraints}
\begin{tabular}{lccccccc}
\hline
Parameter & Truth & CNN+2D & FC+alm & FC+2PCF & FC+WST & Fisher (2PCF) & Likelihood (2PCF) \\
\hline
$\Omega_b$ & 0.0493 & $0.0491 \pm 0.0014$ & $0.0501 \pm 0.0024$ & $0.0502 \pm 0.0032$ & $0.0517 \pm 0.0019$ & $\pm 0.0032$ & $0.0491\pm 0.0032$\\
$\Omega_m$ & 0.315  & $0.3244 \pm 0.0085$ & $0.3235 \pm 0.0129$ & $0.3235 \pm 0.0178$ & $0.3174 \pm 0.0099$ & $\pm 0.0074$ & $0.3000\pm 0.0115$\\
$h$        & 0.674  & $0.6783 \pm 0.0083$ & $0.6704 \pm 0.0163$ & $0.6698 \pm 0.0200$ & $0.6665 \pm 0.0103$ & $\pm 0.0208$ & $0.6727\pm 0.0270$\\
$10^{9}A_s$& 2.100  & $2.0144 \pm 0.0539$ & $2.1127 \pm 0.0638$ & $2.1159 \pm 0.1826$ & $2.0682 \pm 0.0616$ & $\pm 0.0971$ & $2.1057\pm 0.1880$\\
$n_s$      & 0.965  & $0.9711 \pm 0.0094$ & $0.9565 \pm 0.0084$ & $0.9616 \pm 0.0193$ & $0.9722\pm 0.0077$ & $\pm 0.0088$ & $1.0736\pm 0.0160$\\
$\sigma_8$ & 0.811  & $0.8071 \pm 0.0132$ & $0.8123 \pm 0.0170$ & $0.8125 \pm 0.0287$ & $0.8028 \pm 0.0237$ & $\pm 0.0189$ & $0.8189\pm 0.0236$ \\
\hline
\end{tabular}
\end{table*}

Figure~\ref{fig:contour} presents the 1D and 2D posterior distributions obtained from all four models, while Table~\ref{tab:1d_constraints} summarizes the marginalized $1\sigma$ uncertainties for the six cosmological parameters, including, for comparison, the Fisher-matrix forecasts and the results from the full 2PCF likelihood analysis.

The results indicate that all four models recover the fiducial values with reasonable accuracy, though the precision varies among parameters and methods. For $\Omega_b$, all methods produce mean values consistent with the truth. The CNN+2D model achieves the tightest constraint ($\sigma = 0.0014$), outperforming both the Fisher forecast ($\sigma = 0.0032$) and the likelihood analysis ($\sigma = 0.0032$). FC+WST also provides a strong constraint ($\sigma = 0.0019$), comparable to the CNN+2D result. In contrast, FC+alm and FC+2PCF yield larger uncertainties ($\sigma = 0.0024$ and $0.0032$, respectively), reflecting a noticeable reduction in constraining power compared to the density-field-based methods.

For $\Omega_m$, the CNN+2D and FC+WST networks recover values closest to the true parameter (0.315), with uncertainties of 0.0085 and 0.0099, respectively. FC+alm and FC+2PCF yield slightly higher means and larger uncertainties, indicating a broader posterior distribution. The Fisher forecast for the 2PCF shows slightly tighter constraints, while the full likelihood analysis produces a lower central value.

The Hubble parameter $h$ shows consistent recovery across methods. CNN+2D provides the tightest constraint ($0.6783 \pm 0.0083$), whereas FC+alm and FC+2PCF give slightly lower means with broader uncertainties. FC+WST has a slightly lower mean (0.6665) with an intermediate uncertainty of 0.0103.
Predictions for the scalar amplitude $A_s$ demonstrate larger differences. CNN+2D, FC+alm, and FC+WST yield mean values within $\sim 2–4\%$ of the true value, with $1\sigma$ uncertainties ranging from 0.0539 to 0.0638. FC+2PCF produces a noticeably broader posterior ($\pm 0.1826$), highlighting reduced constraining power for this parameter.

For the spectral index $n_s$, all methods recover the true value with good precision, although FC+WST and CNN+2D achieve the smallest uncertainties (0.0077–0.0094). FC+alm and FC+2PCF show slightly larger uncertainties and slightly shifted mean values. The likelihood analysis for 2PCF exhibits a significant upward bias relative to the fiducial value.

Finally, for $\sigma_8$, all methods recover the true value reasonably well. CNN+2D and FC+alm provide tighter constraints ($\pm 0.0132$ and $\pm 0.0170$), while FC+2PCF and FC+WST exhibit broader uncertainties. The Fisher forecast lies between the results of these methods, and the 2PCF likelihood result is slightly higher than the fiducial value.

\begin{figure}[htbp]
    \centering
    \includegraphics[width=0.45\textwidth]{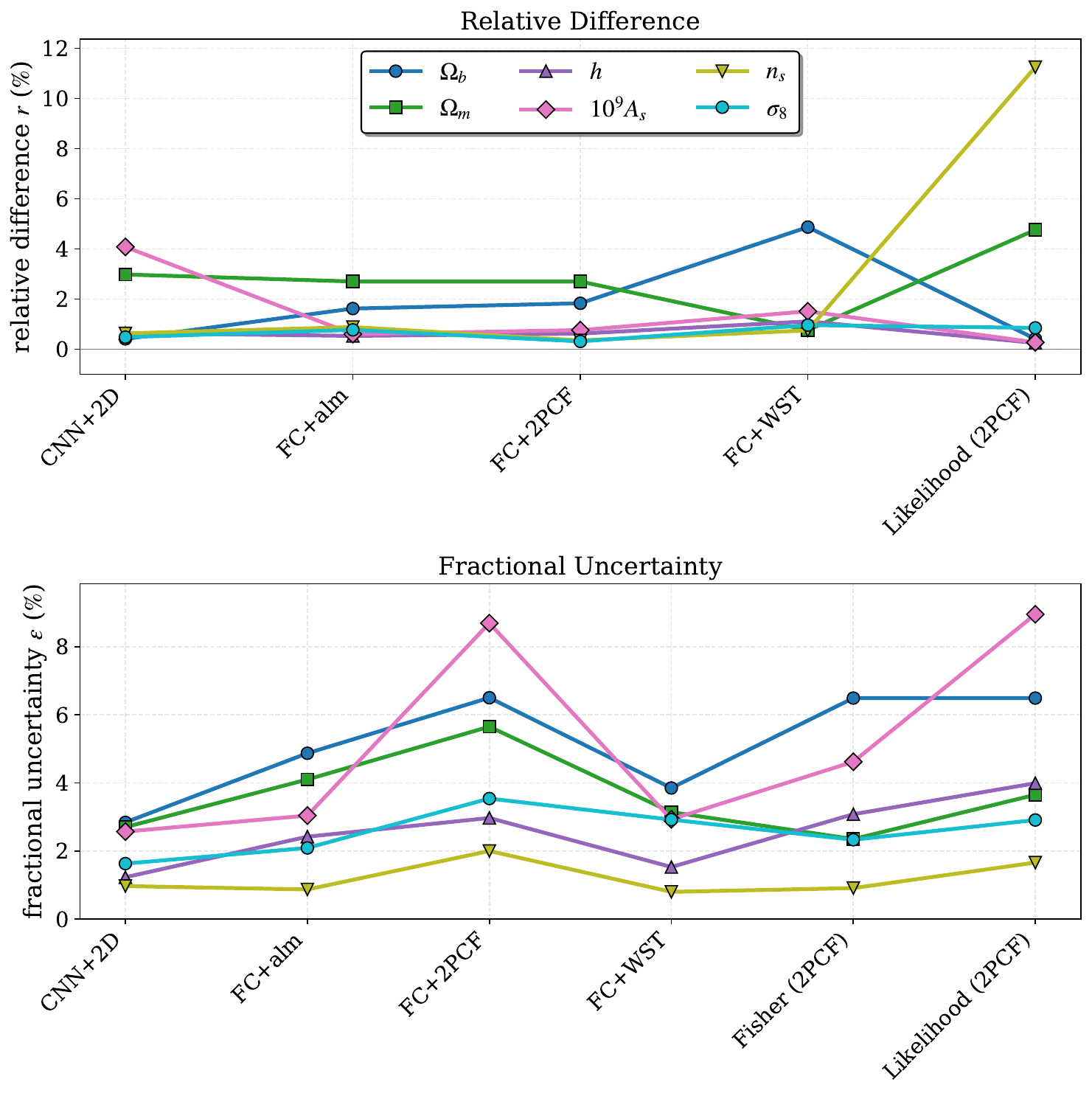}
    \caption{Comparison of relative differences and fractional uncertainties for cosmological parameters across different methods. Upper: the relative bias $r$ (in \%) for our four models and the 2PCF likelihood results. Lower: the fractional uncertainties $\epsilon$ (in \%) for all methods, including 2PCF likelihood results and 2PCF Fisher forecasts. Colors denote different cosmological parameters: $\Omega_b$, $\Omega_m$, $h$, $A_s$, $n_s$, and $\sigma_8$.}
    \label{fig:parameter_bias_error}
\end{figure}

 A more quantitative comparison is shown in Fig.~\ref{fig:parameter_bias_error}, where we define the relative difference, $r = |\theta^{\rm true} - \theta^{\rm pred}|/\theta^{\rm true}$, and the fractional uncertainty, $\epsilon = \sigma/\theta^{\rm true}$, to assess the performance of different methods.

Overall, CNN+2D achieves a favorable balance between low bias and tight constraints. It yields sub-percent-level biases for most parameters, including $\Omega_b$ (0.41\%), $h$ (0.64\%), $n_s$ (0.63\%), and $\sigma_8$ (0.48\%), while maintaining small fractional uncertainties at the $\sim 1\%$--$3\%$ level. In particular, it provides the tightest constraints for $h$ (1.23\%), $n_s$ (0.97\%), and $\sigma_8$ (1.63\%) among all methods.

On the other hand, FC+WST stands out for its exceptionally low bias on certain parameters. It achieves the smallest bias for $\Omega_m$ (0.76\%) and shows competitive performance for $n_s$ (0.75\%) and $\sigma_8$ (0.96\%). Although its uncertainties ($\sim 0.8\%$--$3.9\%$) are slightly larger than those of CNN+2D for some parameters, they remain significantly tighter than those from traditional 2PCF-based approaches.

Overall, CNN+2D and FC+WST exhibit the most competitive performance, offering a strong balance between accuracy and precision across cosmological parameters.


\section{Conclusions}
\label{sec:conclusion}

In this work, we have developed and evaluated a novel framework for cosmological parameter inference from halo lightcone data using two-dimensional convolutional neural networks (CNN+2D). By decomposing the 3D lightcone into narrow redshift slices and projecting each slice onto HEALPix spheres, our approach preserves the approximate translational invariance within each slice while respecting the spherical geometry of the sky. This enables efficient 2D convolutional operations and circumvents the prohibitive computational cost associated with full 3D CNNs.

We applied the CNN+2D method to a set of \texttt{AbacusSummit} halo lightcone mocks covering $0.3 < z < 0.8$ over a $40^\circ \times 40^\circ$ sky patch. For comparison, we also trained fully connected networks on three alternative summary statistics: PCA-compressed $a_{\ell m}$, the 2PCF, and WST coefficients. The predictive performance of all models is first evaluated on a test set of 20 cosmological samples. It is then further quantified using 21 independent realizations of a fiducial cosmology with different random seeds, and benchmarked against traditional 2PCF likelihood analysis and Fisher forecasts.

Our results indicate that CNN+2D consistently extracts physically meaningful cosmological information from the projected density fields, achieving competitive or superior constraints relative to conventional summary statistics. In particular, FC+WST yields the tightest parameter bounds overall, demonstrating the power of non-Gaussian statistical features captured by hierarchical wavelet coefficients. Nevertheless, CNN+2D directly exploits the lightcone’s spatial structure, providing competitive performance efficiently with minimal preprocessing.  Notably, all deep-learning methods outperform the model FC+2PCF in terms of both RMSE and $R^2$. In particular, the CNN+2D, FC+alm, and FC+WST models yield tighter parameter constraints than the likelihood-based 2PCF approach, demonstrating the promising potential of end-to-end field-level feature learning and the added information captured by advanced summary statistics for future LSS surveys.

Looking forward, our 2D CNN methodology can be readily extended to incorporate observational systematics, realistic survey geometries, and galaxy selection effects. Combining CNN-based feature extraction with advanced summary statistics, such as the WST or higher-order correlation functions, may further enhance cosmological constraints. These results establish CNN+2D as a promising tool for analyzing upcoming wide-field galaxy redshift surveys, offering a pathway toward precise and efficient inference of fundamental cosmological parameters from complex three-dimensional datasets.

\begin{acknowledgements}

This study is supported by the National Natural Science Foundation of China (12373005), the National SKA Program of China (2025SKA0160100), the National Natural Science Foundation of China (12473097, 12503008, 12503110), the China Manned Space Project with No. CMS-CSST-2021 (A02, A03, B01), and the Guangdong Basic and Applied Basic Research Foundation (2024A1515012309) and the Fundamental Research Funds for the Central Universities, Sun Yat-sen University(No. 24qnpy122). We wish to acknowledge the Beijing Super Cloud Center (BSCC) and Beijing Beilong Super Cloud Computing Co., Ltd (http://www.blsc.cn/) for providing HPC resources that have significantly contributed to the research results presented in this study.
\end{acknowledgements}

\appendix

\section{Sensitivity of Summary Statistics to Cosmological Parameters}
\label{ssec:stat_comparison}
\begin{figure}[htpb]
    \centering
    \includegraphics[width=0.95\linewidth]{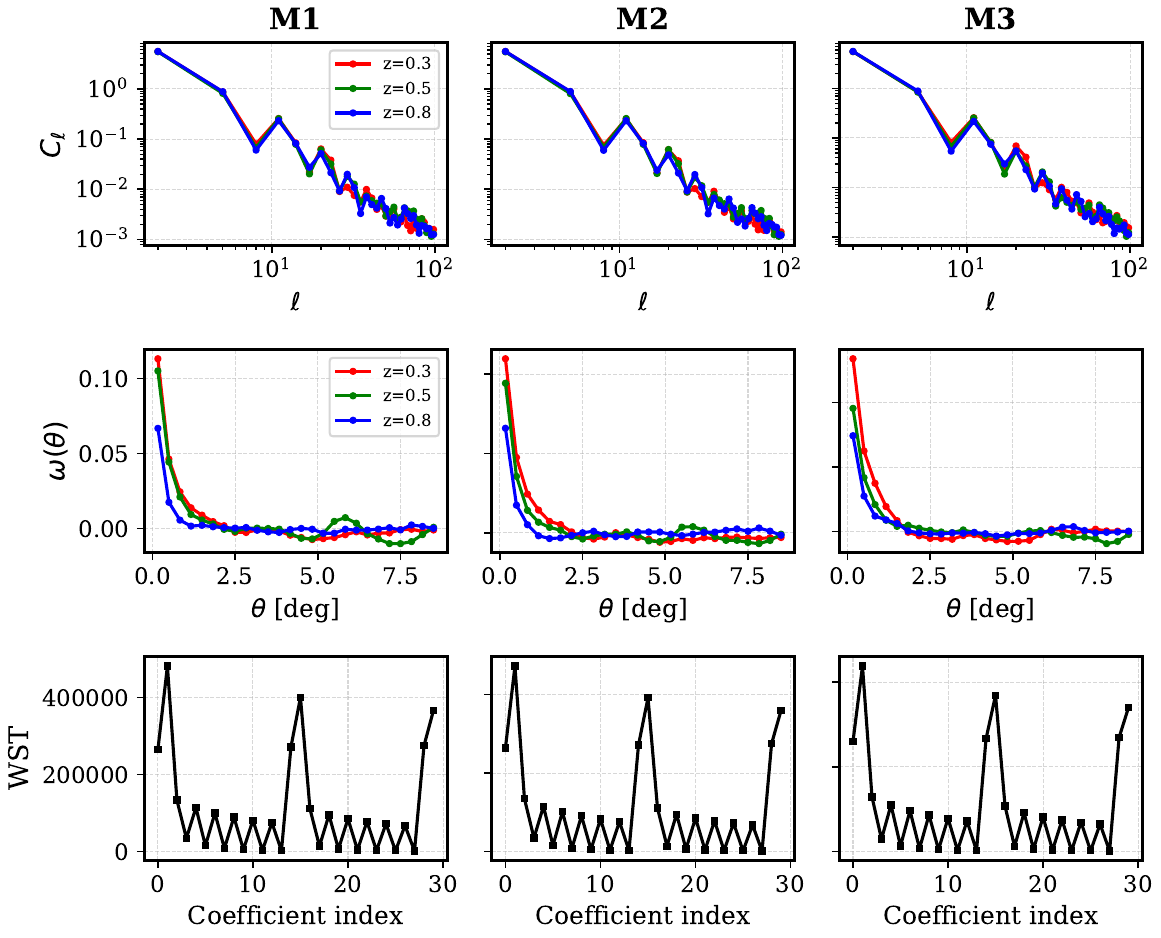}
    \caption{Comparison of the three summary statistics used in this work for three cosmological models. 
    Top: angular power spectrum $C_\ell$ derived from the spherical harmonic coefficients $a_{\ell m}$ in the three representative redshift bins. 
    Middle: 2PCF for the three models in the three redshift bins.
    Bottom: a subset of WST coefficients for one sub-volume, ordered by scale $j$ and orientation $l$. 
    The WST is computed from the 3D density field without redshift binning.}
    \label{fig:npcfwst}
\end{figure}

Figure~\ref{fig:npcfwst} compares the three summary statistics used in this work for three cosmological models (M1, M2, M3) drawn from the \texttt{AbacusSummit} suite, where M3 denotes the cosmological parameter set $\{\Omega_b=0.0391, \Omega_m=0.2485, h=0.7426, 10^9A_s=2.07, n_s=0.9538, w_0=-1.076, w_a=-0.395, \sigma_8=0.8387\}$. The models share identical initial random seeds but differ in cosmological parameters. Therefore, any differences in the statistics arise purely from the change in cosmology. The three models span a representative range of parameters, for example $\Omega_m=0.2485$–$0.3578$ and $\sigma_8=0.7297$–$0.8825$.

The top panel shows the angular power spectra $C_\ell$ derived from the spherical harmonic coefficients $a_{\ell m}$ in three representative redshift bins. Differences between cosmological models are visible across a broad range of multipoles and become more pronounced at smaller angular scales (large $\ell$), reflecting the sensitivity of the power spectrum to nonlinear clustering.

The middle panel presents the angular two-point correlation function $\omega(\theta)$ measured in the same redshift bins. Similar to the power spectrum, the correlation amplitude increases toward lower redshift. However, the cosmological dependence is most visible at small angular separations ($\theta \lesssim 2^\circ$), where nonlinear clustering enhances model differences, while the large-scale signal is relatively similar between models.

The bottom panel shows a subset of the WST coefficients computed from the full 3D density field of a representative sub-volume. Unlike the previous statistics, the WST is applied to the entire lightcone without binning in redshift, allowing it to capture non-Gaussian structures and LoS evolution simultaneously. The coefficients encode hierarchical multi-scale information, and subtle variations between cosmological models can already be seen at different wavelet scales and orientations.

Overall, the three statistics probe different aspects of the halo distribution. The $C_\ell$ and $\omega(\theta)$ quantities  primarily capture two-point clustering information in individual redshift slices, while the WST coefficients encode higher-order, multi-scale structures in the full 3D lightcone. Their differing sensitivities to cosmological parameters and redshift evolution motivate the comparative analysis with a CNN presented in Sect.~\ref{sec:results}.

\section{Jackknife Covariance Estimation for 2PCF}
\label{app:jackknife}

To estimate a robust covariance matrix for the 2PCF measurements, we employ the jackknife resampling method \citep{Norberg:2008tg}. For each redshift slice, we divide the $40^\circ\times40^\circ$ sky area into $N_{\rm sub}=16$ non-overlapping $10^\circ\times10^\circ$ sub-regions. The 2PCF is measured 16 times, each time omitting one sub-region, yielding 16 jackknife realizations of the 2PCF for that slice.

\begin{figure}
    \centering
    \includegraphics[width=0.8\linewidth]{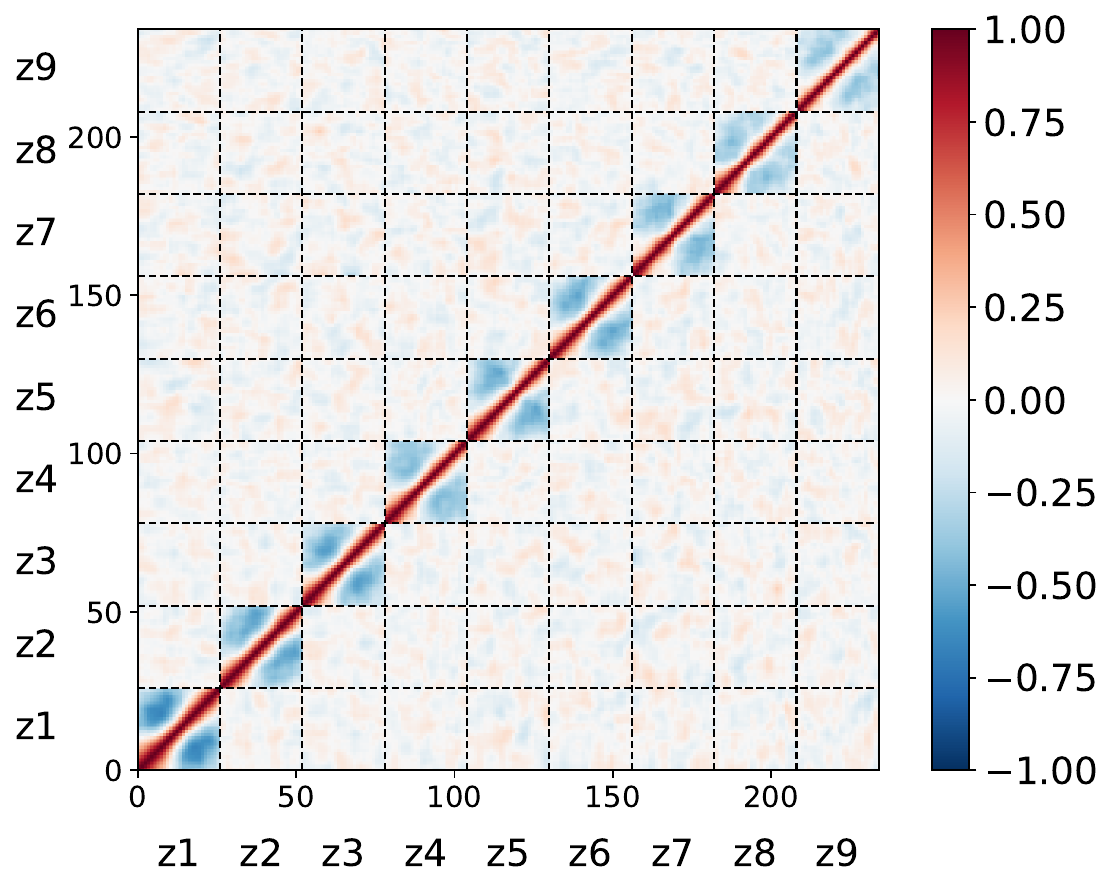}
    \caption{Correlation matrix of all $9 \times 26$ 2PCF measurements evaluated at the fiducial cosmology. Black dashed lines indicate the boundaries between the nine different redshift bins.}
    \label{fig:corre_matrix}
\end{figure}

The full data vector for each jackknife realization is constructed by concatenating the 2PCF measurements from all nine redshift slices, resulting in a vector of length $9\times26 = 234$. The covariance matrix is then estimated from the scatter among these concatenated jackknife samples:

\begin{equation}
\label{eq:jackknife}
C_{ij} = \frac{N_{\rm sub}-1}{N_{\rm sub}} \sum_{a=1}^{N_{\rm sub}} (\xi_i^a - \bar{\xi}_i)(\xi_j^a - \bar{\xi}_j),
\end{equation}

where $\xi_i^a$ is the $i$-th element of the concatenated data vector from the $a$-th jackknife realization, and $\bar{\xi}_i$ is the mean over all jackknife samples for that element. This approach naturally captures both the correlations within each redshift slice and the cross-correlations between different slices, as the jackknife resampling is performed consistently across all slices. Figure~\ref{fig:corre_matrix} shows the resulting correlation matrix, revealing strong correlations within the same redshift slice (the diagonal blocks) and weaker correlations between different slices (off-diagonal blocks), as expected.

\section{2PCF Emulator Validation}
\label{app:loo}

\begin{figure}
    \centering
    \includegraphics[width=0.8\linewidth]{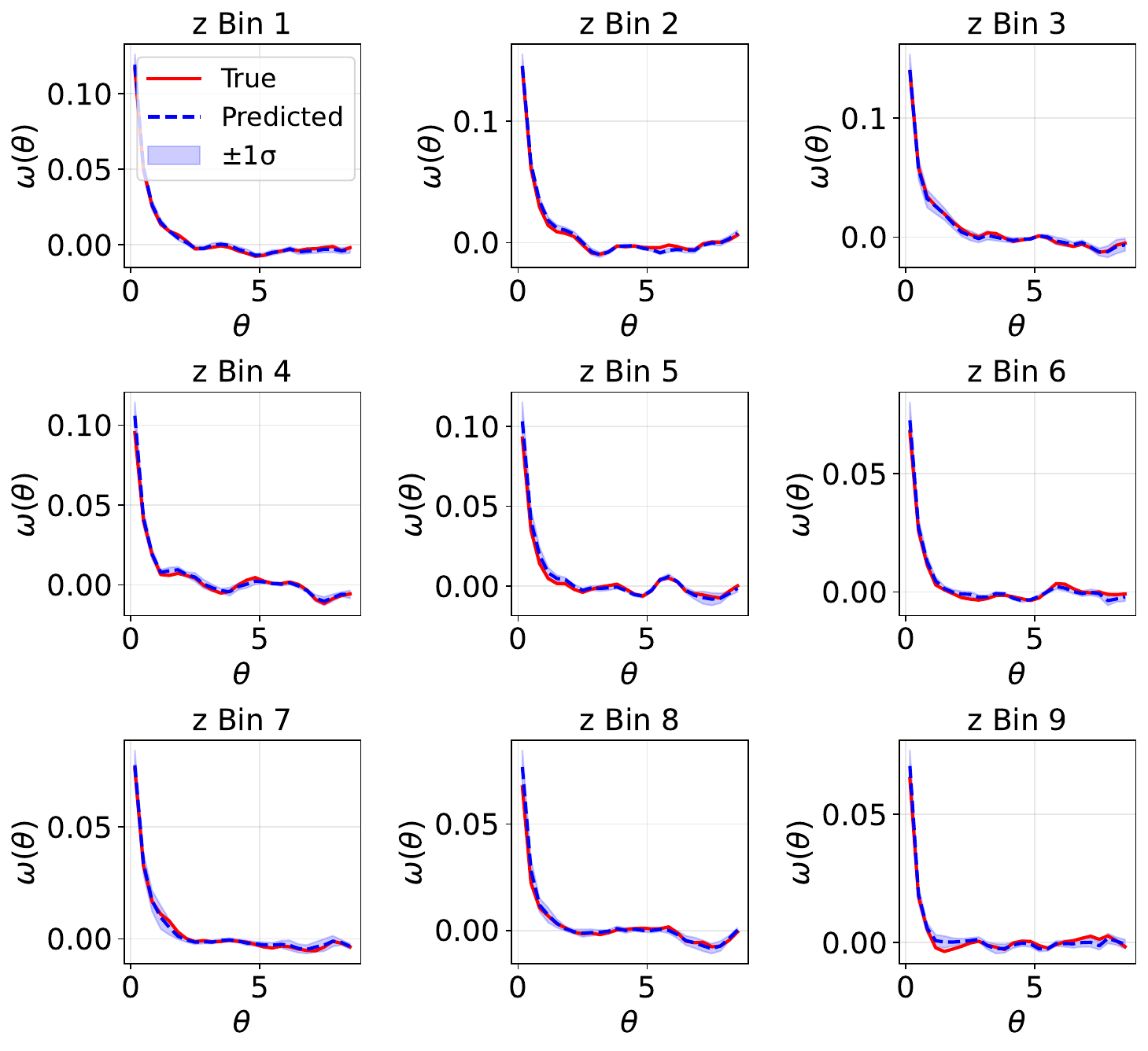}
    \caption{Leave-One-Out validation results for a representative sample. The red solid line shows the true 2PCF values, while the blue dashed line represents the emulator predictions. The shaded region indicates the $\pm 1\sigma$ uncertainty range of the GP emulator.}
    \label{fig:loo}
\end{figure}

To obtain the model prediction $m(\theta)$ for any point in parameter space, we train a 2PCF emulator using GPR regression on the 52 training cosmologies. The GP models the smooth variation of the 2PCF as a function of the six cosmological parameters. We validate the emulator's accuracy using Leave-One-Out (LOO) cross-validation, where a single training cosmology is held out and predicted by a GP trained on the remaining 51. Figure~\ref{fig:loo} shows the LOO validation results for a representative sample, demonstrating that the emulator accurately reproduces the true 2PCF measurements across all redshift slices.

\bibliography{apssamp}

\end{document}